\documentclass[fleqn,11pt]{article}
\usepackage{amsmath,amssymb}
\usepackage{xcolor}
\usepackage{verbatim}

\newtheorem{stat}{Property}

\newtheorem{Mark}{Comment}

\newcommand{\Mod}[1]{\mathbf{#1}}
\newcommand{\ModTilde}[1]{\tilde{\mathbf{#1}}}
\newcommand{\Fig}[3]{%
\begin{center}
\parbox{8cm}{%
\refstepcounter{figure}\includegraphics[width=8cm,height=#2cm]{#1} \noindent Fig. \thefigure:\quad
#3}\end{center}}
\newcounter{strochka}

\newcounter{spisok}
\usepackage{amsfonts,amssymb,cite}
\usepackage{graphicx}

\def\noi{\noindent}

\newcommand{\Title}[1]{\noi {{\Large\bf #1}}\\[1ex]}

\newcommand{\Author}[2]{\noi{\bf #1}\\[2ex]\noi{\normalsize\it #2}\\}

\newcommand{\Abstract}[1]{\vskip 2mm \begin{center}
        \parbox{16.4cm}{\small\noi #1} \end{center}\medskip}
\newcommand{\foom}[1]{\protect\footnotemark[#1]}

\def\nqq{\hspace*{-2em}}

\def\Jl#1#2{#1 {\bf #2},\ }

\def\ApJ#1 {\Jl{Astroph. J.}{#1}}
\def\CQG#1 {\Jl{Class. Quantum Grav.}{#1}}
\def\DAN#1 {\Jl{Dokl. AN SSSR}{#1}}
\def\GC#1 {\Jl{Grav. Cosmol.}{#1}}
\def\GRG#1 {\Jl{Gen. Rel. Grav.}{#1}}
\def\JETF#1 {\Jl{Zh. Eksp. Teor. Fiz.}{#1}}
\def\JETP#1 {\Jl{Sov. Phys. JETP}{#1}}
\def\JHEP#1 {\Jl{JHEP}{#1}}
\def\JMP#1 {\Jl{J. Math. Phys.}{#1}}
\def\NPB#1 {\Jl{Nucl. Phys. B}{#1}}
\def\NP#1 {\Jl{Nucl. Phys.}{#1}}
\def\PLA#1 {\Jl{Phys. Lett. A}{#1}}
\def\PLB#1 {\Jl{Phys. Lett. B}{#1}}
\def\PRD#1 {\Jl{Phys. Rev. D}{#1}}
\def\PRL#1 {\Jl{Phys. Rev. Lett.}{#1}}

\def\lal{&&\nqq {}}

\def\beq{\begin{equation}}
\def\eeq{\end{equation}}
\def\bear{\begin{eqnarray}}
\def\bearr{\begin{eqnarray} \lal}
\def\ear{\end{eqnarray}}
\def\earn{\nonumber \end{eqnarray}}

\begin{document}
\thispagestyle{empty}
\twocolumn[

\vspace{1cm}

\Title{Formation of supermassive nuclei of Black holes in the early Universe by the mechanism of scalar-gravitational instability. I. Local picture.   \foom 1}

\Author{Yu. G. Ignat'ev}
    {Institute of Physics, Kazan Federal University, Kremlyovskaya str., 16A, Kazan, 420008, Russia}

\Abstract
 {Based on the formulated and proven similarity properties of cosmological models based on a statistical system of degenerate scalarly charged fermions, as well as the previously identified mechanism of scalar-gravitational instability of cosmological models, a numerical-analytical study of the formation of supermassive black hole nuclei in the early Universe was carried out. A mathematical model of the evolution of spherical perturbations is constructed, on the basis of which the main regularities of the process of evolution of collapsing masses and the dependence of the parameters of forming black holes on the fundamental parameters of the cosmological model and the wavelength of gravitational perturbations are revealed. In this case, the mass loss of the black hole due to quantum evaporation is taken into account. A stable tendency for the early formation of supermassive black hole nuclei in the class of cosmological models under study is shown, and a close connection between the growth of masses of spherical perturbations and the nature of the singular points of these models is shown.\\[8pt]
}
\bigskip

] 

\section{Introduction}
The article is devoted to the still unsolved theoretical problem of early formation\footnote{When the age of the Universe is less than one billion years, $z\gtrsim 6$.} supermassive black hole seeds, SSBH,\footnote{\textbf{S}eeds of \textbf{ S}upermassive \textbf{B}lack \textbf{H}oles -- \textbf{SSBH}.} $\sim10^9M_\odot$ at quasar centers (see \cite{SMBH1e} -- \cite{Soliton})
\begin{equation}\label{M_nc}
m_{ssbh}\sim 10^4\div 10^6 M_\odot\approx 10^{42}\div10^{44}m_{\mathrm{pl}}.
\end{equation}
The standard mechanisms of hydrodynamic and gas gravitational instability (\cite{Lifshitz}, \cite{Land_Field}) are not capable of providing such a rapid growth of unstable masses. This problem is considered in sufficient detail in the articles \cite{Yu_Unst_GC22_1} -- \cite{Yu_Unst_GC23_1}. In these papers, in particular, it was noted that such an early formation of SSBH, when scalar fields can have a significant impact on cosmological processes, leads to the need to take into account the interaction of scalarly charged matter with scalar fields in the theory of gravi\-ta\-tional instability.

On this path, \cite{GC_21_1} -- \cite{GC_21_2} in the so-called hard WKB approximation, the instability of short - wavelength longitudinal modes of plane scalar - gravitational perturbations in a cosmological system of degenerate scalar-charged fermions with the\\ Higgs interaction was discovered. The discovered \emph{scalar-gravitational instability} was further inves\-ti\-gated in \cite{Yu_Unst_GC22_1} -- \cite{Yu_Unst_GC23_1}: in \cite{Yu_Unst_GC22_2} -- \cite{Yu_Unst_GC22_3} -- a one-field model with a canonical \cite{Yu_Unst_GC22_1}, \cite{Yu_Unst_GC23_1} -- two-field model with canonical and phantom scalar fields. A mathematical model of flat short - wavelength scalar - gravitational perturbations in a cosmological system of degenerate scalar-charged fermions with the Higgs interaction was studied in \cite{Yu_Unst_PPJ22_1} -- \cite{Yu_Unst_RPJ22_2}. The scalar-gravitational instability of the cosmological medium is fun\-da\-men\-tally different from the hydrodynamic instability of E.M. Lifshitz (see \cite{Lifshitz}, \cite{Land_Field}), which arises as a result of the Jeans gravitational instability mechanism: this instability, in contrast to the short-wavelength gravitational instability of a cosmological fluid, arises in the zero WKB approximation, due to which the amplitude of perturbations can grow with time not according to the power law, but according to the exponential law.

The physical nature of the scalar-gravitational instability lies in the combination of the collective scalar interaction of scalarly charged
fermions\footnote{having a number of unique features (see \cite{Yu_Unst_GC23_2})} and gravitational interaction. The results of previous works indicate that scalarly charged matter plays the role of a catalyst for the process of gravitational instability, binding the scalar field to itself, but at the same time making an insignificant contribution to the energy balance of the system.

The exponential growth of perturbations quickly brings the process to a nonlinear stage, at which, due to macroscopic isotropy, interacting plane per\-tur\-ba\-tions must be reformed into spherical ones \cite{Yu_Unst_GC22_3}. Further evolution of spherical disturbances is studied in detail in \cite{YuTMF_23}. \cite{Yu_Unst_GC23_1}, \cite{YuTMF_23} shows the fundamental possibility of the formation of Black Holes with parameters \eqref{M_nc} in the early Universe using the scalar-gravitational instability mechanism. In particular, the similarity property of self - gravi\-tating systems of scalarly charged particles is proved in \cite{YuTMF_23}, which can be used to extend the numerical results to systems with similar sets of fundamental constants.

In this part of the work, we will apply the results obtained to determine the parameters of objects that arise as a result of the development of scalar-gravitational instability. In particular, we will apply the recently discovered similarity properties of the studied cosmological models \cite{arXiv_23}, which allow, firstly, to carry out numerical simulation of processes in the region of physically significant parameter values, which was practically inaccessible due to excessive time costs, and, secondly, to analytically extend obtained numerical results for a large range of model parameters. Thanks to these properties, it was possible to achieve significant progress in the study of the designated problem. In the second part of the work, we plan to investigate the large-scale consequences of the SSBH formation process, including finding upper bounds on the parameters of these objects and clarifying the cosmological history after the SSBH formation process is\\ completed.

\section{Cosmological system of scalarly charged fermions}
Let us briefly formulate the main provisions of the macroscopic theory\footnote{In \cite{TMF_21} it is shown how this theory is obtained from microscopic dynamics.} for a cosmological model based on a one-component degenerate statistical system of scalarly charged fermions and a scalar Higgs field $\Phi$.
%
\subsection{General Model Equations}
The Lagrange function $L_s$ of the scalar Higgs field is\footnote{Here and below, Latin letters run over $\overline{1,4}$, Greek -- $\overline{1,3}$.}
\begin{eqnarray} \label{L_s}
L_s=\frac{1}{16\pi}(g^{ik} \Phi_{,i} \Phi_{,k} -2V(\Phi)),
\end{eqnarray}
where
\begin{eqnarray}
\label{Higgs}
V(\Phi)=-\frac{\alpha}{4} \left(\Phi^{2} -\frac{m_s^{2}}{\alpha}\right)^{2}
\end{eqnarray}
is the potential energy of the scalar field, $\alpha$ is the self-action constant, $m_s$ is the mass of the quanta of the scalar field. Energy-momentum tensors of scalar fields relative to the Lagrange function \eqref{L_s}, $S^i_{k}$, and \emph{equilibrium} statistical system, $T^i_{k}$, are
\begin{eqnarray}\label{T_s}
S^i_{k} =\frac{1}{16\pi}\bigl(2\Phi^{,i}\Phi_{,k}- \delta^i_k\Phi_{,j} \Phi^{,j}+2V(\Phi)\delta^i_k \bigr);
\end{eqnarray}
\begin{equation}\label{T_p}
T^i_{k}=(\varepsilon_p+p_p)u^i u_k-\delta^i_k p_p,
\end{equation}
where $u^i$ is the velocity vector of the statistical system and $\varepsilon_p$, $p_p$ are the energy density and pressure of the statistical system.

Einstein's equations for the ``scalar field + par\-ticles'' system are:
\begin{equation}\label{Eq_Einst_G}
R^i_k-\frac{1}{2}\delta^i_k R=8\pi (T^i_k+S^i_k) + \delta^i_k \Lambda_0,
\end{equation}
where $\Lambda_0$ is the \emph{seed} value of the cosmological constant, related to its \emph{observed} value $\Lambda$ obtained by removing the constant term in the potential energy by the relation:
\begin{equation}\label{lambda0->Lambda}
\Lambda=\Lambda_0-\frac{1}{4}\frac{m_s^4}{\alpha}.
\end{equation}

The macroscopic consequences of the kinetic theory are the transport equations \cite{TMF_21}, including the law of conservation of some vector current corresponding to the microscopic law of conservation of some fundamental charge %
\begin{equation}\label{1}
\nabla_i q n^i=0,
\end{equation}
as well as the laws of conservation of energy - momentum of the statistical system:
\begin{equation}\label{2}
\nabla _{k} T_{p}^{ik} -\sigma\nabla^{i} \Phi =0,
\end{equation}
where $\sigma$ is the density of scalar charges with respect to the field $\Phi$ \cite{TMF_21}. The \eqref{2} equations are equivalent to the equations of ideal hydrodynamics
\begin{eqnarray}\label{2a}
(\varepsilon_p+p_p)u^i_{~,k}u^k=(g^{ik}-u^iu^k)(p_{p,k}+\sigma\Phi_{,k});\\
\label{2b}
\nabla_k[(\varepsilon_p+p_p)u^k]=u^k(p_{p,k}+\sigma\Phi_{,k}),
\end{eqnarray}
and the fundamental charge conservation laws \eqref{1}:
\begin{equation}\label{2c}
\nabla_k \rho u^k=0,
\end{equation}
where $\rho\equiv q n$ is \emph{the kinematic density of the scalar charge}.

Macroscopic scalars for a one-component statistical system of degenerate fermions take the form:
\begin{eqnarray}
\label{2_3c}
n=\frac{1}{\pi^2}\pi_f^3;\; p_p  =\displaystyle \frac{e^4\Phi^4}{24\pi^2}(F_2(\psi)-4F_1(\psi));\\
\label{2_3a_2}
 \sigma=\frac{e^4 \Phi^3}{2\pi^2}F_1(\psi);   \; \varepsilon_p=\frac{e^4 \Phi^4}{8\pi^2}F_2(\psi),
\end{eqnarray}
where $\pi_f$ is the Fermi momentum, $\sigma$ is the density of scalar charges $e$ and
\begin{equation}\label{psi}
\psi=\frac{\pi_f}{|e\Phi|}
\end{equation}
and the functions $F_1(\psi)$ and $F_2(\psi)$ are introduced:
\begin{eqnarray}\label{F_1}
F_1(\psi)=\psi\sqrt{1+\psi^2}-\ln(\psi+\sqrt{1+\psi^2});\nonumber\\
\label{F_2}
\!\!\!F_2(\psi)=\psi\sqrt{1+\psi^2}(1+2\psi^2)-\ln(\psi+\sqrt{1+\psi^2}).\nonumber
\end{eqnarray}
In this case, the equations of the scalar field for a system of scalarly charged degenerate fermions are obtained as a consequence of the transport equations
\begin{eqnarray}\label{Box(Phi)=sigma_z}
\Box \Phi + m_s^2\Phi-\alpha\Phi^3 =-8\pi\sigma\equiv-\frac{4e^4\Phi^3}{\pi} F_1(\psi).
\end{eqnarray}
Thus, the mathematical model $\Mod{M}$ of a self-gravitating system of scalarly charged fermions consists of the Einstein equations \eqref{Eq_Einst_G}, the hydrodynamic equations \eqref{2} and the scalar field equation \eqref{Box(Phi)=sigma_z} together with definitions of sources: the energy-momentum tensors of the scalar field \eqref{T_s}, the fermionic component, \eqref{T_p}, and the scalar charge density, \eqref{2_3a_2}, as well as the fermion energy density \eqref{2_3c} and their pressure \eqref {2_3a_2}. As can be seen from the equations of this system and the definition of its coefficients, the solutions of the Cauchy problem for this system of equations for given \emph{fundamental parameters} $\mathbf{P}=[\alpha,m_s,e,\Lambda_0]$ are completely determined by the corresponding initial conditions with respect to the metric functions $g_{ik}(x^j)$, the potential $\Phi(x^j)$, the velocity vector $u^i(x^j)$, and the Fermi momentum $\pi_f$.

In \cite{YuTMF_23}, \cite{arXiv_23} the similarity property of the con\-si\-dered dynamical system is proved.

\begin{stat}\label{stat1}
\hskip -6pt\textbf{.} $\blacksquare$ The complete system of equations of the mathematical model $\Mod{M}$ is invariant with respect to simultaneous scaling transformations of fundamental parameters $\Mod{P}$ \eqref{trans_param}, coordinates and Fermi momentum \eqref{trans_x} of the mathematical model
\begin{eqnarray}\label{trans_param}
\mathcal{S}_k(\Mod{M}): &  \alpha=k^2\tilde{\alpha},\; m_s=k\tilde{m}_s;& \nonumber\\
& e=\sqrt{k}\tilde{e};\;\Lambda= k^2\tilde\Lambda;&\\
\label{trans_x}
& x^i= k^{-1}\tilde{x}^i,\quad \pi_f= \sqrt{k}\tilde{\pi}_f, & .
\end{eqnarray}
($k=\mathrm{Const}>0$) that is, under scaling tran\-s\-for\-ma\-tions \eqref{trans_param} -- \eqref{trans_x} and the corresponding tran\-s\-for\-mation of the initial conditions, the solutions of the equations of the original model $\Mod{M}$ and the scaling transformed model $\ModTilde{M}$ coincide:
\begin{eqnarray}\label{trans_eqs}
\!\!\!\!\Phi(x)=\tilde{\Phi}(\tilde{x}); g_{ik}(x)= \tilde{g}_{ik}(\tilde{x}); u^i(x)= \tilde{u}^i(\tilde{x}).\ \blacksquare
\end{eqnarray}
\end{stat}
The \textbf{\ref{stat1}} similarity property of a mathematical model allows extending a solution with a given set of fundamental parameters to the case of other values of fundamental parameters. This is practically im\-por\-tant in the numerical integration of the model equations in the case of very small or very large values of the parameters.

Under scaling transformations \eqref{trans_param} -- \eqref{trans_x} both parts of the equations \eqref{Eq_Einst_G}, \eqref{2} and \eqref{Box(Phi)=sigma_z} are multiplied by $k^2$, and the introduced above, scalars and tensors change according to the laws:
\begin{eqnarray}\label{trans_scalar}
\psi=\tilde{\psi},\; \sigma= k^2\tilde{\sigma};\;V(\Phi)= k^2 \tilde{V}(\tilde{\Phi});\nonumber\\
p_p= k^2 \tilde{p}_p;\; \varepsilon_p= k^2 \tilde{\varepsilon}_p;\nonumber\\
S^i_k= k^2 \tilde{S}^i_k;\quad T^i_k= k^2 \tilde{T}^i_k.
\end{eqnarray}
\subsection{Cosmological model equations}
In the case of a spatially flat Friedmann metric
\begin{eqnarray}\label{ds_0}
ds_0^2=dt^2-a^2(t)(dx^2+dy^2+dz^2)\equiv \\
dt^2-a^2(t)[dr^2+r^2(d\theta^2+\sin^2\theta d\varphi^2)],\nonumber
\end{eqnarray}
and homogeneous isotropic distribution of matter $\Phi=\Phi(t);\; \pi_f=\pi_f(t);\; u^i=\delta^i_4$ the energy-momentum tensor of a scalar field takes the form of the energy-momentum tensor of an ideal isotropic fluid:
\begin{equation} \label{MET_s}
S^{ik} =(\varepsilon_s +p_{s} )u^{i} u^{k} -p_s g^{ik} ,
\end{equation}
at that
\begin{eqnarray}\label{Es-Ps}
\varepsilon_s=\frac{1}{8\pi}\biggl(\frac{\dot{\Phi}^2}{2}+V(\Phi)\biggr);\\
p_{s}=\frac{1}{8\pi}\biggl(\frac{\dot{\Phi}^2}{2}-V(\Phi)\biggr).
\end{eqnarray}
The material equations \eqref{2} -- \eqref{2a} integrate exactly \cite{TMF_21}:
\begin{equation}\label{aP0}
 a\pi_f=\mathrm{Const},
\end{equation}
and the function $\psi$ \eqref{psi} is completely defined in terms of the functions $a(t)$ and $\Phi(t)$:
\begin{equation}\label{psi(t)}
\psi=\frac{\pi_0}{|e\Phi|}\mathrm{e}^{-\xi}, \quad (\pi_0=\pi_f(0)),
\end{equation}
where we have passed to the new variable $\xi(t)$
\begin{equation}\label{a-xi}
\xi=\ln a,
\end{equation}
opining here and in the future:
\begin{equation}\label{xi(0)}
\xi(0)=0.
\end{equation}

In this case, the system of equations \eqref{Eq_Einst_G}, \eqref{2} and \eqref{Box(Phi)=sigma_z} reduces to the autonomous dynamic system $\Mod{M}$ \cite{TMF_21}, \cite{Ignat_GC21} :
\begin{eqnarray}\label{dot_xi-dot_Phi}
\dot{\xi}=H; \; \dot{\Phi}=Z;\;\\
\label{dH/dt_0}
\dot{H}=- \frac{Z^2}{2}-\frac{4}{3\pi}e_z^4\Phi^4\psi^3\sqrt{1+\psi^2};\\
\label{dZ/dt}
\dot{Z}=-3HZ-m_s^2\Phi +\Phi^3\biggl(\alpha-\frac{4e^4}{\pi}F_1(\psi)\biggr),
\end{eqnarray}
where $H(t)$ is the Hubble parameter. The Einstein equation for the $^4_4$ component is the first integral of the \eqref{dot_xi-dot_Phi} -- \eqref{dZ/dt} system:
\begin{eqnarray}\label{Surf_Einst1_0}
3H^2-\Lambda-\frac{Z^2}{2}-\frac{m_s^2\Phi^2}{2}\nonumber\\
+\frac{\alpha\Phi^4}{4}-\frac{e^4\Phi^4}{\pi}F_2(\psi)=0.
\end{eqnarray}
The \eqref{Surf_Einst1_0} equation defines some three-dimensional hypersurface $\mathbb{S}_3$ in the four-dimensional arithmetic phase space of the dynamical system $\mathbb{S}_3\subset \mathbb{R}_4=\{\xi,H,\Phi ,Z\}$, on which all phase trajectories of the dynamical system lie, in other words, each specific cosmological model corresponds to one line on this hypersurface. Following \cite{Ignat_Dima_GC20}, we will call $\mathbb{S}_3$ an \emph{Einstein-Higgs hypersurface}. The \eqref{Surf_Einst1_0} equation can be considered as a definition of the initial value of the Hubble parameter $H(0)\equiv H_0$ under given initial conditions for the rest of the dynamical variables. The dynamical system \eqref{dot_xi-dot_Phi} -- \eqref{dZ/dt} is invariant with respect to time translations $t\to t+t_0$ due to its autonomy, and this allows us to choose \eqref{xi(0)} as the initial condition ($\xi_0=0$). Thus, two quantities remain free: $\Phi_0$ and $Z_0$. Since, due to \eqref{Surf_Einst1_0}, the Einstein-Higgs hypersurface is symmetric with respect to the $H=0$ plane, then for given $\Phi_0,Z_0$ we have two symmetric solutions for the initial value of the Hubble parameter $H^\pm_0=\pm H_0$. One of these solutions, $H^+_0$, corresponds to starting from the expansion state, the second, $H^-_0$, from the con\-trac\-tion state.

Thus, under the condition when choosing the sign of the initial value of the Hubble parameter, only two initial values remain free: $\Phi_0$ and $Z_0$, which we will also specify by an ordered list
\begin{eqnarray}\label{Inits}
\mathbf{I}=[\Phi_0,Z_0,\pm1], & (\Phi_0=\Phi(0),Z_0=Z(0)).
\end{eqnarray}
Taking into account the exact integral \eqref{aP0}, the initial value of the Fermi momentum $\pi_0$ will also be assumed to be the fundamental parameter of the cosmological model, setting further the fundamental parameters of the model $\mathbf{M}$ ordered list \cite{TMF_21}:
\begin{eqnarray}\label{Par}
\mathbf{P}=[[\alpha,m_s,e,\pi_0],\Lambda].
\end{eqnarray}
\subsection{Singular points of a dynamical system}
The following property of the $\Mod{M}$ model (dynamical system) \cite{arXiv_23} takes place.
\begin{stat}\label{stat3}$\blacksquare$
Coordinates of eigenpoints of the dynamical system of the cosmological model $\Mod{M}$ at $H\not\equiv 0$ in the subspace $\mathbb{R}_3\equiv\{H,\Phi,Z\}\subset \mathbb{R }_4$, as well as the eigenvalues of the characteristic matrix, coincide with the coordinates of the eigenpoints and the eigenvalues of the characteristic matrix of the vacuum-field cosmological model. $\blacksquare$
\end{stat}
The coordinates of these six singular points in the subspace $\mathbb{R}_3=\{H,\Phi,Z\}$ of the phase space of the dynamical system are (see \cite{arXiv_23} -- \cite{YuKokh_TMF}):
\begin{eqnarray}\label{M_0}
M^\pm_0=\biggl[\pm\sqrt{\frac{\Lambda}{3}},0,0\biggr];\\
\label{M_pm}
M^\pm_\pm=\biggl[\pm\sqrt{\frac{\Lambda_0}{3}},\pm\frac{m_s}{\sqrt{\alpha}},0\biggr],
\end{eqnarray}
and the nonzero eigenvalues of the characteristic matrix of the system at these singular points are:
\begin{eqnarray}\label{eigen_val_0}
M^\pm_0:& \!\!\!\!
\left|\begin{array}{ll}
\lambda_2=& \pm\sqrt{\displaystyle \frac{\Lambda}{3}},\\[8pt]
\lambda_{3,4}=& \mp\frac{1}{2}\sqrt{3\Lambda} \pm \frac{1}{2}\sqrt{3\Lambda-4m^2_s};\\
\end{array}\right.\\
\label{eigen_val_pm}
M^\pm_\pm:&
\left|\begin{array}{ll}
\lambda_2=& \displaystyle \pm\sqrt{\frac{\Lambda_0}{3}},\\[8pt]
\lambda_{3,4}=& \mp\frac{1}{2}\sqrt{3\Lambda_0} \pm \frac{1}{2}\sqrt{3\Lambda_0+8m^2_s}.\\
\end{array}\right.
\end{eqnarray}

\begin{Mark}\hspace{-6pt}\textbf{.}\label{mark1}
The investigated cosmological model $\Mod{M}$, based on a system of scalarly charged particles, in contrast to the vacuum-field cosmological model $\Mod{M_0}$, in which the vacuum scalar field is the only source of gravity, always has a singular state in the final moment of cosmological time $t_0$, such that
\begin{equation}\label{t_0}
a(t_0)=0,
\end{equation}
(see, for example, \cite{TMF_21}, \cite{Ignat_GC21}). With an arbitrary choice of consistent initial conditions \eqref{Inits} at time $t=0$ this moment of time most often turns out to be negative.
\end{Mark}
\begin{Mark}\hspace{-6pt}\textbf{.}\label{mark2}
The results of numerical simulation show that the cosmological model $\Mod{M}$ for the time $\Delta t\gtrsim (\Lambda_0)^{-1/2}$ goes from the ultrarelativistic stage of expansion to the inflationary one, which, in turn, can have two stages corresponding to different values of the Hubble parameter - first the larger $H_1=\sqrt{\Lambda_0/3}$ and then the smaller $H_2=\sqrt{\Lambda/3}$.
\end{Mark}

\begin{Mark}\hspace{-6pt}\textbf{.}\label{mark3}
Near unstable singular points of the $\Mod{M}$ dynamical system, the similarity symmetry of cosmological models can be broken \cite{arXiv_23}. Violation occurs precisely because of instability. In order to avoid this, when integrating numerically, it suffices to slightly shift the initial conditions \eqref{Inits} so that they do not coincide with the coordinates of the unstable singular point.
\end{Mark}
\subsection{Similarity Properties of\newline a Mathematical Model}

Consider two cosmological models: $\mathbf{M}$ with fundamental parameters $\mathbf{P}$ and initial conditions $\mathbf{I}$
and a similar model $\tilde{\mathbf{M}}$ with fundamental parameters $\tilde{\mathbf{P}}$ and initial conditions $\tilde{\mathbf{I}}$ --
\begin{eqnarray}
\label{Inits_tilde}
\tilde{\mathbf{I}}=\biggl[\Phi_0,\frac{1}{k}Z_0\biggl];\\
\label{Par_tilde}
\tilde{\mathbf{P}}=\biggl[\biggl[\frac{\alpha}{k^2},\frac{m_s}{k},\frac{e}{\sqrt{k}},\frac{\pi_0}{\sqrt{k}}\biggr],\frac{\Lambda}{k^2}\biggr].
\end{eqnarray}
Functions $f(t)=\tilde{f}(\tilde{t})$ that are invariant under the similarity transformation \eqref{trans_param} -- \eqref{trans_x} are transformed according to the rules:
\begin{equation}\label{f=f}
\tilde{f}(\tilde{t})=f\biggl(\frac{\tilde{t}}{k}\biggr).
\end{equation}
Let the solutions of the dynamical system \eqref{dot_xi-dot_Phi} -- \eqref{dZ/dt}, \eqref{Surf_Einst1_0} for the model $\mathbf{M}$ \eqref{Inits} -- \eqref{Par} be
\[\mathbf{S}(t)=[\xi(t),H(t),\Phi(t),Z(t)].\]
Then, according to the property \textbf{\ref{stat1}}, the solutions of the corresponding equations for such a model $\tilde{\mathbf{M}}$ \eqref{Inits_tilde} -- \eqref{Par_tilde} are
\begin{eqnarray}\label{Sol_tilde}
\tilde{\mathbf{S}}(t)=[\tilde{\xi}(t),\tilde{H}(t),\tilde{\Phi}(t),\tilde{Z}(t)]\equiv \nonumber \\
\biggl[\xi\biggl(\frac{t}{k}\biggl),\frac{1}{k}H\biggl(\frac{t}{k}\biggl),\Phi\biggl(\frac{t}{k}\biggl),\frac{1}{k}Z\biggl(\frac{t}{k}\biggl)\biggr].
\end{eqnarray}

Note that under scaling transformations \eqref{trans_param} -- \eqref{trans_x} the singular state time $t_0$ of the model is transformed according to the transformation law of the time coordinate $t$, including its negative values\footnote{which is fully confirmed by numerical integration}:
\begin{equation}\label{t0->t0}
\tilde{t}_0=kt_0.
\end{equation}

In \cite{arXiv_23} the following similarity property of the considered dynamical system $\Mod{M}$ is proved.

\begin{stat}\label{stat2}
\hskip -6pt\textbf{.} $\blacksquare$Under scaling transformations \eqref{trans_param} -- \eqref{trans_x} the eigenvalues of the characteristic matrix of the dynamical system \eqref{dot_xi-dot_Phi} -- \eqref{dZ/dt} are transformed according to the law
\begin{equation}\label{tilde_lambda=}
\tilde{\lambda}=\frac{\lambda}{k}.
\end{equation}
In this case, the coordinates of singular points are transformed, as well as arbitrary coordinates of the phase trajectory of the dynamical system, i.e., according to the law \eqref{Sol_tilde}.

Due to the proportionality of the eigenvalues of such models, the nature of singular points is an invariant property of similarity. Thus, the behavior of similar cosmological models is also similar.
$\blacksquare$
\end{stat}

The values of fundamental parameters (model $\tilde{M}$)\footnote{Such values correspond, for example, to the scales of field models of the SU(5) type, are physically realizable.}:
\begin{eqnarray}\label{Phys_Par}
\alpha\lesssim10^{-8}; m\lesssim10^{-4};e\lesssim10^{-2}.
\end{eqnarray}
For such small values of the fundamental parameters, the standard methods of numerical integration of a nonlinear system of dynamic equations do not allow extending calculations to sufficiently large values of the time $t\gtrsim 10^3$ (see, for example, \cite{TMF_21}, \cite{Ignat_GC21}. Making scale transformation with a similarity coefficient of the order of $k=10^4$, we will pass to the cosmological model $M$ with parameters $\alpha=m=e=1$, which is already amenable to numerical simulation up to much larger time values $t \sim 10^8$ Using the results of integrating the $M$ model according to the scaling transformation rules, we thereby extend the results for the original $\tilde{M}$ model up to times of the order of $t\gtrsim 10^7$.

Another physically important circumstance is that when passing from the cosmological model $M$ under study to a similar cosmological model $\tilde{M}$ with similarity coefficient $k\sim 10^4\div10^5$, we expand by the same number of times time interval of cosmological evolution, passing to the times $kt\gg t_{pl}$, at which the quantum-field consideration of the cosmological model is not required, but rather its classical description. Making the reverse transition from the classical model $\tilde{M}$ to the similar model $M$ with the similarity coefficient $k^{-1}$, we obtain the classical model at times comparable to the Planck ones. This model, however, by no means claims any physical meaning. It serves only as a \emph{computational model} similar to the classical cosmological model under study for large evolution times $t\gg t_{pl}$.

\begin{Mark}\hspace{-6pt}\textbf{.} \label{mark4}
It is easy to see that the invariant cosmological acceleration $\Omega$
\begin{equation}\label{Omega}
\Omega=\frac{\ddot{a}}{\dot{a}^2}\equiv 1+\frac{\dot{H}}{H^2}
\end{equation}
is also invariant with respect to the similarity trans\-for\-mation:
\begin{equation}\label{tilde_Omega}
\tilde{\Omega}(\tilde{t})=\Omega(t).
\end{equation}
Since the barotropic coefficient $\kappa=p/\varepsilon$, which determines the equation of state, is linearly related to the cosmological acceleration, an important con\-clu\-sion follows from this:
the equation of state is invariant with respect to the similarity trans\-for\-mation -- similar cosmological models have similar equations of state:
\begin{equation}\label{tilde_kappa}
\tilde{\kappa}(\tilde{t})=\kappa(t).
\end{equation}
\end{Mark}
\section{Cosmological evolution\newline of perturbations}
Let us briefly formulate the main results of the theory of short-wave plane perturbations in the cosmological medium of single-sorted scalar-charged degenerate fermions (see \cite{Yu_Unst_GC22_2} -- \cite{Yu_Unst_GC22_3}, mathematical model of perturbations \cite{Yu_Unst_PPJ22_1}--\cite{Yu_Unst_RPJ22_2}).
\subsection{Expansion in plane\newline shortwave perturbations}
We write the metric with gravitational perturbations for the case of purely longitudinal perturbations of the metric \eqref{ds_0} in the standard form (see \cite{Land_Field}):
\begin{eqnarray}
\label{metric_pert}
ds^2=ds^2_0-a^2(\eta)h_{\alpha\beta}dx^\alpha dx^\beta,
\end{eqnarray}
where $ds_0$ is the unperturbed Friedmann metric \eqref{ds_0} in conformally flat form
\begin{eqnarray}
\label{ds0eta}
ds^2_0=a^2(\eta)(d\eta^2-dx^2-dy^2-dz^2),
\end{eqnarray}
and the time variable $\eta$ is related to the cosmological time $t$ by the relation
\begin{equation}\label{t=eta}
t=\int a(\eta)d\eta.
\end{equation}
The non-zero components of the perturbations of the metric are \footnote{for definiteness, the wave vector is directed along the $Oz$ axis.}:
\begin{eqnarray}\label{nz1}
 h_{11}=h_{22} =\frac{1}{3}[\lambda(t)+\frac{1}{3}\mu(t)]\mathrm{e}^{inz};\nonumber\\
\label{nz2}
h_{33}=\frac{1}{3}[-2\lambda(t)+\mu(t)]\mathrm{e}^{inz}.
\end{eqnarray}
Matter is completely determined by two scalar functions -- $\Phi(z,\eta)$ and $\pi_{z}(z,\eta)$, as well as by the velocity vector $u^i(z,\eta)$. Let us expand these functions into a series in terms of the smallness of perturbations with respect to the corresponding functions against the background of the Friedmann metric \eqref{ds0eta}:
\begin{eqnarray}
\Phi(z,\eta)=\Phi(\eta)+\delta\Phi(\eta)\mathrm{e}^{inz};\nonumber\\
\label{dF-drho-du}
\pi_f(z,t)=\pi_f(\eta)(1+\delta(\eta)\mathrm{e}^{inz});\nonumber\\
\sigma(z,\eta)= \sigma(\eta)+\delta\sigma(\eta)\mathrm{e}^{inz};\\
u^i=\frac{1}{a}\delta^i_4+\delta^i_3 v(\eta)\mathrm{e}^{inz},\nonumber
\end{eqnarray}
where $\delta\Phi(\eta)$, $\delta(\eta)$, $\delta\sigma(\eta)$, and $v(\eta)$ are functions of the first order of smallness compared to their unperturbed values.

In accordance with the WKB method, we re\-pre\-sent the per\-tur\-bation functions $f(\eta)$ in the form
\begin{eqnarray}\label{petrurb}
\displaystyle f(\mathbf{r},t,\mathbf{n})=\tilde{f}(\mathbf{n},t)\mathrm{e}^{i\mathbf{nr}+\int u(\mathbf{n},t)dt/a(t)}=\\
\label{Eiconal}
\tilde{f}(\eta) \cdot \mathrm{e}^{i\int u(\eta)d\eta}; \quad (|u\eta|\sim |n\eta| \gg 1),
\end{eqnarray}
where $\tilde{f}(\eta)$ and $u(\eta)$ are functions of the per\-tur\-bation amplitude and eikonal that vary slightly along with the scale factor.

Note that the condition for the applicability of the WKB appro\-xi\-mation \eqref{Eiconal} in terms of the cosmological time $t$ \eqref{t=eta} means
\begin{equation}\label{apr_wkb}
r_n(t)\equiv n\eta(t)\gg1\Rightarrow n\int\limits_{t_0}^t \mathrm{e}^{-\xi}dt\gg1,
\end{equation}
therefore, as calculations show, the WKB appro\-xi\-ma\-tion is applicable even for small values of $n\lesssim1$ \cite{Yu_Unst_GC22_2}.

Expanding the field equation \eqref{Box(Phi)=sigma_z} in a Taylor series in terms of the smallness of the perturbations, we obtain equations for the perturbation of the first-order scalar field $\delta\Phi$:
\begin{eqnarray}\label{Eq_dPhi}
\delta\Phi''+2\frac{a'}{a}\delta\Phi'+\bigl[n^2+a^2(m^2_s-3\alpha\Phi^2)\bigr]\delta\Phi\nonumber\\
+\frac{1}{2}\Phi'\mu'=-8\pi a^2\delta\sigma,
\end{eqnarray}
Introducing a new variable for gravitational per\-tur\-ba\-tions
\begin{equation}\label{nu}
\nu=\lambda+\mu
\end{equation}
and expanding the Einstein equations \eqref{Eq_Einst_G} into a Taylor series in perturbation orders, we obtain the following independent equations for gravitational perturbations of first-order perturbations $\nu,\lambda$\\ (primed denotes conformal time derivatives $\eta$):
\begin{eqnarray}
\label{34}
v=\frac{in}{8\pi a^3(\varepsilon+p)_p}\biggl(\delta\Phi\Phi' +\frac{1}{3}(\lambda'+\mu')\biggr);\\
\label{Eq_lambda}
\lambda''+2\frac{a'}{a}\lambda'-\frac{1}{3}n^2\nu=0;\\
\label{Eq_nu}
\nu''+2\frac{a'}{a}\nu'+\frac{1}{3}n^2\nu+3\delta\Phi'\Phi'\nonumber\\
-3 a^2[\Phi\delta\Phi(m^2_s-\alpha\Phi^2)-8\pi\delta p_p]=0.
\end{eqnarray}
Wherein, the equation \eqref{34} determines the longi\-tu\-dinal velocity of disturbances. Thus, the evolution of plane perturbations with wavenumber $n$ is comp\-le\-tely determined by the system of three ordinary linear differential equations \eqref{Eq_dPhi}, \eqref{Eq_lambda} and \eqref{Eq_nu}.

\subsection{WKB approximation and dispersion equation}
In the zero WKB approximation \eqref{Eiconal}, the equations for the perturbation amplitudes \eqref{Eq_dPhi}, \eqref{Eq_lambda}, and \eqref{Eq_nu} take the form of a linear homogeneous system of algebraic equations
\begin{eqnarray}\label{WKB0-sys}
\!\!\!
\left[
\begin{array}{ccc}
n^2-u^2+\gamma_{11} & 0 & n^2\gamma_{13}\\
0 & u^2 & 0 \\
\gamma_{31} & 0 & n^2\gamma_{33}-u^2\\
\end{array}
\right]\cdot
\left[
\begin{array}{c}
\delta\Phi\\ \lambda \\ \nu\\
\end{array}
\right]=0,\nonumber
\end{eqnarray}
where the coefficients $\gamma_{\alpha\beta}$ are:
\begin{eqnarray}
\label{gamma_ik}
\gamma_{11}\equiv a^2(m^2_s-3\alpha\Phi^2+8\pi S_\Phi);\nonumber\\
\gamma_{13}\equiv \frac{e^4_z\Phi^3\psi^2}{6\pi^2\varepsilon^\delta_p\sqrt{1+\psi^2_z}};\; \gamma_{33}\equiv \frac{1}{3}+\frac{p^\delta_p}{\varepsilon^\delta_p};\\
\gamma_{31}\equiv -3a^2[\Phi(m^2_s-\alpha\Phi^2)-8\pi P^\Phi].\nonumber
\end{eqnarray}
The coefficients of the theory of scalar-gravitational instability \cite{Yu_Unst_RPJ22_2} $S_\Phi$, $\varepsilon^\delta_p$, $p^\delta_p$ included in the formulas \eqref{gamma_ik} are determined in terms of the functions of the background solution $ a(t)$, $\Phi(t)$, and $\psi(t)$ by very cumbersome formulas (see, for example, \cite{Yu_Unst_RPJ22_2}).

A necessary and sufficient condition for the nontrivial solvability of the system of equations \eqref{WKB0-sys} is that the determinant of the matrix of this system is equal to zero, which gives the necessary \emph{dispersion equation} on eikonal functions $u(t)$ of perturbations. In this case, two zero modes $u^\pm_{(0)}=0$ are immediately distinguished, corresponding to perturbations of the $\lambda$ metric (see \cite{Yu_Unst_GC22_1}), which are eliminated by admissible transformations. Four non-zero oscillation modes $u^\pm_{(\pm)}$ corresponding to $\delta\Phi$ scalar field perturbations and $\nu$ metric perturbations are defined by the dispersion equation
\begin{eqnarray}\label{A_gamma=0}
\!\!\mathrm{Det}(\bar{\mathbf{A}})=\left|\begin{array}{ll}
n^2-u^2+\gamma_{11}  & n^2\gamma_{13}\\[12pt]
\gamma_{31} &  n^2\gamma_{33}-u^2\\
\end{array}\right|=0,
\end{eqnarray}
having solutions:
\begin{eqnarray}
\label{u_pm}
u^\pm_\pm=\pm\frac{1}{\sqrt{2}}\biggl(n^2(1+\gamma_{33})+\gamma_{11}\nonumber\\
\pm\frac{1}{2}\sqrt{[n^2(1-\gamma_{33})+\gamma_{11}]^2+4\gamma_{13}\gamma_{31}}\biggr)^{1/2},
\end{eqnarray}
where the upper signs correspond to the signs before the external radical, the lower ones correspond to the sign before the internal one. \eqref{u_pm} solutions satisfy the following relations:
\begin{eqnarray}\label{u-=-u+}
u^-_\pm=-u^+_\pm;\; u^+_-u^+_+=u^-_-u^-_+.
\end{eqnarray}
Depending on the signs of the expressions under the radicals in \eqref{u_pm}, there are only 4 types of perturbations depending on the value of $\gamma_{ik}$: the 1st type represents a superposition of a pair of standing growing and damping modes and a pair of undamped\footnote{Weak amplitude attenuation occurs only due to the geometric factor $a(t)$.} of delayed and advanced waves, the 2nd type represents two pairs of undamped waves (leading and retarded) with different frequencies, the 3rd type is two pairs of traveling waves (leading and delayed) with different frequencies, having damped and growing modes, the 4th type represents pairs of damped and growing with time with different decrements/increments of standing waves. Since the coefficients in the expression \eqref{u_pm} are functions of time -- $\gamma_{ik}(t)$, the listed types of perturbations can transform one into another over time. As a result, in the course of cosmological evolution, a rather complex picture of the alternation of the stages of wave oscillations and the stages of instability, determined by the presence of the imaginary part of the eikonal \cite{Yu_Unst_GC22_1}, can be obtained.

Returning to the eikonal perturbation rep\-re\-sen\-tation \eqref{Eiconal} and passing from the conformal time $\eta$ to the cosmological time $t$ \eqref{t=eta}, we write the perturbation amplitudes
\begin{equation}\label{Sol_f}
f(t,z)=\sum\limits_{\pm}^{\pm}\tilde{f}^\pm_\pm \mathrm{e}^{i(\pm nz \pm\int \omega^\pm_\pm dt)}\mathrm{e}^{-\int\gamma^\pm_\pm dt},
\end{equation}
where
\begin{eqnarray}\label{g,o}
\omega^\pm_\pm=\mathrm{e}^{-\xi(t)}\mathrm{Re}\left(u^\pm_\pm\right)\equiv \mathrm{e}^{-\xi(t)}\bar{\omega}^\pm_\pm;\nonumber\\
\gamma^\pm_\pm=-\mathrm{e}^{-\xi(t)}\mathrm{Im}\left(u^\pm_\pm\right)\equiv \mathrm{e}^{-\xi(t)}\bar{\gamma}^\pm_\pm,
\end{eqnarray}
Growing oscillation modes correspond to the insta\-bi\-lity of the homogeneous unperturbed state of the cosmological model. As we noted above, this mode is associated with $\{\delta\Phi,\nu\}$ disturbances, so the instability, if it exists, is essentially \emph{scalar-gravitational} in nature. As we noted above, the nature of the scalar - gravitational instability is fun\-da\-men\-tally different from the hydrodynamic instability \cite{Lifshitz} - it arises in the WKB order zero and develops according to the exponential law, like the usual Lyapunov instability. Note that the coefficients $\gamma_{ik}$ \eqref{gamma_ik} and, together with them, the frequencies $\omega^\pm_\pm,\gamma^\pm_\pm$ depend on the cosmological time via the background functions $\xi( t),\Phi(t)$
\[\omega^\pm_\pm(t)\equiv\omega^\pm_\pm(\xi(t),\Phi(t));\; \gamma^\pm_\pm(t)\equiv\gamma^\pm_\pm(\xi(t),\Phi(t)),\]
and also on the fundamental parameters of the $\mathbf{P}$ model. Note that at the moment of transition to the standing wave regime, the amplitude of the wave propagating in one of the directions grows exponentially rapidly, while the amplitude of the wave propagating in the opposite direction also decreases exponentially. As a result, only one osci\-l\-lation mode survives.

\emph{The perturbation amplitude growth factor} for the growing perturbation mode at time $t$ is given by
\begin{equation}\label{chi}
\chi(t)=\int_{t_1}^t \gamma(t)dt,
\end{equation}
where $t_1$ is the initial moment of instability occurrence. Let $t_2$ be the end time of the unstable phase, so that for $t>t_2$ $\gamma(t)=0$. Thus, during the development of instability on the interval $\Delta t=t_2-t_1$, the perturbation amplitude is fixed at $\tilde{f}^+(t)\exp(\chi_\infty)$, where
\[\chi_\infty=\int_{t_1}^{t_2} \gamma(t)dt.\]

Consider the following example of the fun\-da\-mental parameters of the cosmological model:
 \begin{equation}\label{F_param0}
\mathbf{P_0}= [[1,1,1,1],0.02],
 \end{equation}
moreover, here and in what follows \textbf{everywhere}, for simplicity, we assume that the initial conditions are unified:
 \begin{equation}\label{Inits0}
 \mathbf{I_0:}\; [1,0,1].
 \end{equation}
On Fig. \ref{fig1} plots of functions $\omega^\pm_-(\xi,\Phi=1)$ and $\gamma^\pm_-(\xi,\Phi=1)$ for perturbation modes $u^\pm_-$ , dashed lines $\omega^\pm_-$, solid lines $\gamma^-_-$, dash-dotted lines $\gamma^+_-$, and in Fig. \ref{fig2} plots of functions $\omega^\pm_+(\xi,\Phi=1)$ and $\gamma^\pm_+(\xi,\Phi=1)$ for perturbation modes $u^\pm_+$, black dashed lines -- $\omega^\pm_+$, solid lines -- $\gamma^-_+$, dash-dotted lines -- $\gamma^+_+-$.

Commenting on the graphs in Fig. \ref{fig1}--\ref{fig2}, we note, firstly, that each figure shows the frequencies and damping increments (growth increments) for oscillations propagating in opposite directions.\\ Secondly, on the left side of Fig. \ref{fig1} we see exactly the waves propagating in opposite directions with a short decay/rise period. In the right part of this figure, at $\xi\gtrsim0$, the oscillations stop ($\omega^\pm_-=0$) and turn into two standing waves, the amplitude of one of which increases, the other one decreases, i.e., a scalar-gravitational instability develops. As the results of numerical simulation show, with a further increase in $\xi$, the damping decrement (growth increment) of oscillations is fixed at a constant value. At the same time, the oscillation modes $u^\pm_+$ do not form standing waves, but represent a pair of waves with the opposite direction of propagation, of which one wave undergoes a short-term attenuation, and the second wave, on the contrary, gains. Thus, out of four oscillation modes, only one gives growing standing waves, i.e., scalar-gravitational instability.

\Fig{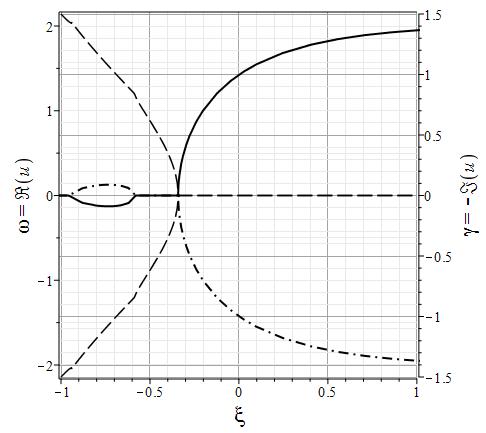}{6}{\label{fig1}Dependence of the eikonal functions of the $u^\pm_-(\xi)$ modes for the model with the parameters \eqref{F_param0}, $n=1$.}
\Fig{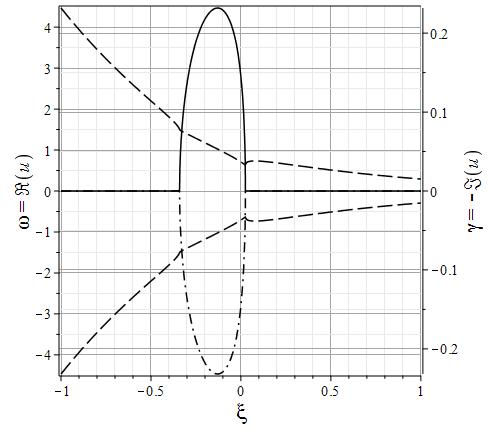}{6}{\label{fig2}Dependence of the eikonal functions of the $u^\pm_+(\xi)$ modes for the model with the parameters \eqref{F_param0}, $n=1$.}

\subsection{Similarity of perturbation theory quantities}
Due to the similarity property \ref{stat1}, \eqref{Eiconal} and \eqref{t=eta} under transformations \eqref{trans_param}--\eqref{trans_x} the perturbation theory values are transformed as follows (we set $f=f(t ),\tilde{f}=\tilde{f}(\tilde{t})$):
\begin{eqnarray}
\eta = k^{-1}\tilde{\eta};\; n=k\tilde{n};\; u=k\tilde{u};\; \gamma_{11}=k^2\tilde{\gamma}_{11};\nonumber \\
\gamma_{13}=k^2\tilde{\gamma}_{13};\;\gamma_{31}=k^2\tilde{\gamma}_{31};\; \gamma_{33}=\tilde{\gamma}_{33};\nonumber \\
\omega^\pm_\pm=k\tilde{\omega}^\pm_\pm;\; \gamma^\pm_\pm=k\tilde{\gamma}^\pm_\pm;\quad \chi_\infty=\tilde{\chi}_\infty.
\end{eqnarray}
Thus, although the oscillation growth rate $\gamma^\pm_\pm$ of two similar models differs, the finite disturbance growth factor $\chi_\infty$ remains invariant.

\begin{Mark}\hspace{-6pt}\textbf{.} \label{mark5}Obviously, \eqref{apr_wkb} is invariant under scaling \eqref{trans_param}--\eqref{trans_x} %
\begin{equation}\label{R->R}
{\displaystyle\tilde{r}_{\tilde{n}}}(\tilde{\mathbf{P}},\tilde{\mathbf{I}},\tilde{t})=r_n(\mathbf{P},\mathbf{I},kt),
\end{equation}
therefore, the applicability condition for the WKB approximation is invariant with respect to the similarity transformation. If, however, the wave number $n$ is fixed under scaling transformations, that is, we consider the image of perturbations with the conservation of the wave number, then instead of \eqref{R->R}, we obtain the transformation law:
\begin{equation}\label{R->kR}
\tilde{r}_n(\tilde{\mathbf{P}},\tilde{\mathbf{I}},\tilde{t})=k r_n(\mathbf{P},\mathbf{I},kt).
\end{equation}
Therefore, even if the condition \eqref{apr_wkb} is weakly satisfied for values of the fundamental parameters of order one, i.e., $r_n\lesssim1$, under the transformation \eqref{trans_param}--\eqref{trans_x} with the similarity coefficient $k\ sim10^{4}$ to the real values of the fundamental parameters and the conservation of the wave number $n$, the WKB applicability condition \eqref{apr_wkb} begins to be satisfied with a large margin $\tilde{r}_n\gg1$.
\end{Mark}

\subsection{Examples of numerical modelling of the evolution of plane perturbations}
For numerical modelling of the evolution of short-wavelength perturbations, it is necessary to find the time dependence of the coefficients \eqref{gamma_ik} $\gamma_{ik}(t)$, for which it is necessary to first determine by numerical integration the basic functions of the cosmological model $\xi(t)$ and $\Phi (t)$. To illustrate the theoretical model of perturbation instability, we present two examples of numerical modelling of the cosmological evolution of instability for the parameters of the \eqref{F_param0} model, referring the Reader to the works \cite{Yu_Unst_GC22_1} -- \cite{Yu_Unst_RPJ22_2}, which present detailed results of numerical modelling, both for one-field and two-field models of scalar interaction.

\noindent \textbf{Example 1. Aperiodic instability: $\mathbf{P}=\mathbf{P}_0$}. On Fig. \ref{fig3} and Fig. \ref{fig4} graphs of frequency evolution (dashed lines) and disturbance growth increments (solid color) for oscillation modes $u^-_+, u^-_-$ -- gray and $u^+_+, u^ +_-$ -- black. Perturbations $u^-_+, u^+_+$ represent a pair of waves propagating in opposite directions, one of which is damping, the other growing. Short-term instability phases, during which the wave amplitude changes, occur twice.
\Fig{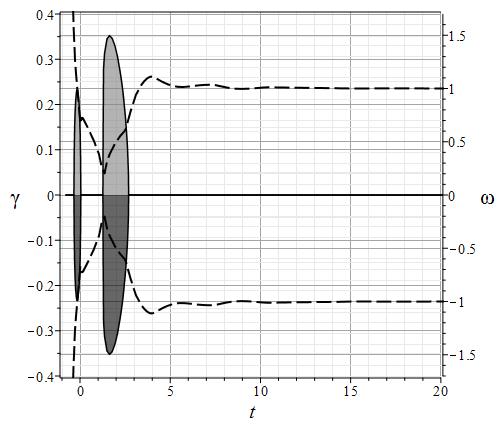}{6}{\label{fig3}Evolution of $u^\pm_+(t)$ for a model with parameters \eqref{F_param0}, $n=1$.}

In contrast to these perturbation modes, in the $u^-_-, u^+_-$ modes, perturbations periodically change the nature of the phase of standing oscillations, during which the amplitude of perturbations increases or decreases, and are replaced by phases of free propagation of oscillations. At the same time, the amplitude of growth of disturbances $\chi(t)$ in the $u^-_-$ mode eventually grows, while in the $u^+_+$ mode it decreases (Fig. \ref{fig5}).

The finite growth of the perturbation amplitude \eqref{chi} is not large, $\exp(\chi_\infty)\approx 1.5$, so the example has a demonstration character.
\Fig{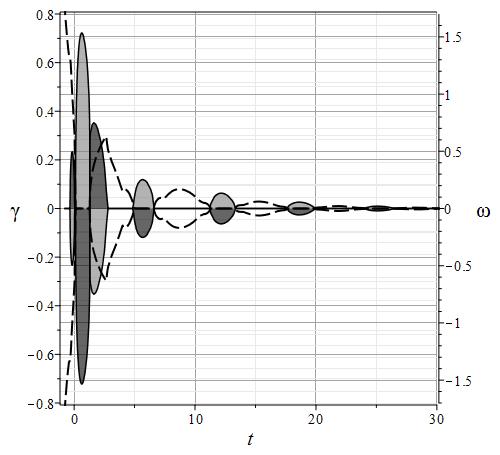}{6}{\label{fig4}Evolution of $u^\pm_-(t)$ for a model with parameters \eqref{F_param0}, $n=1$.}

\Fig{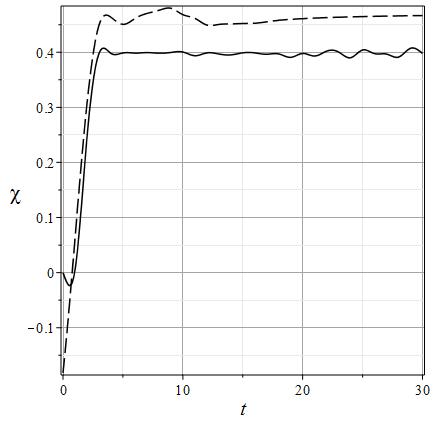}{6}{\label{fig5}Evolution of $\chi(t)$ \eqref{chi} for the modes $u^-_+(\xi)$ (dashed line) and $u^-_-(\xi)$ (solid line) for the model with parameters \eqref{F_param0}, $n=1$.}

\noindent \textbf{Example 2. Broadband Unstability}:\footnote{This example is taken from \cite{Yu_Unst_GC22_1}, the graphs are slightly reformatted.}
\begin{equation}\label{F_param1}
\mathbf{P}=\mathbf{P}_1=[[1,1,10^{-7},0.1],0.003].
\end{equation}
In this case, the $u^\pm_-(\xi)$ perturbation modes quickly turn into standing waves of constant amplitude, while the $u^\pm_+(\xi)$ perturbations give one damped and one constantly growing mode:
\begin{equation}\label{shir_moda}
\gamma^\pm_+(\xi)\approx \mp 1.4;\;  \quad \chi(t)\approx \mp 1.4\cdot t;\; (t\gtrsim10).
\end{equation}

Therefore, the perturbation modes $u^\pm_-(\xi)$ in this case can grow exponentially quickly to very large values, thereby violating the linear approximation condition:
\begin{equation}\label{exp}
f(t)\backsimeq \tilde{f}(t)\mathrm{e}^{1.4t}.
\end{equation}
So, for example, at $\tilde{f}\sim 10^{-6}$ the perturbations grow up to a value on the order of unity in the time $t\sim 25$. Here, however, one must remember that the parameters $\mathbf{P_1}$ \eqref{F_param1} are physically unacceptable. The real values of the parameters are obtained from \eqref{F_param1} with a similarity coefficient of the order of $k\sim 10^{-4}\div10^{-5}$.

Performing scaling transformations with such a value of the coefficient $k$ according to the rules \eqref{trans_param} -- \eqref{trans_x}, we get instead of \eqref{exp}
\[f(t)\backsimeq \tilde{f}(t)\exp\bigl[1.4\cdot(10^{-5}\div10^{-4})t\bigr].\]
But in this case, the time for perturbations to reach the nonlinear stage also increases by 4-5 orders of magnitude, which is already $10^4\div10^5$ Planck times\footnote{That is, it occurs already at the classical expansion stage.}.
\Fig{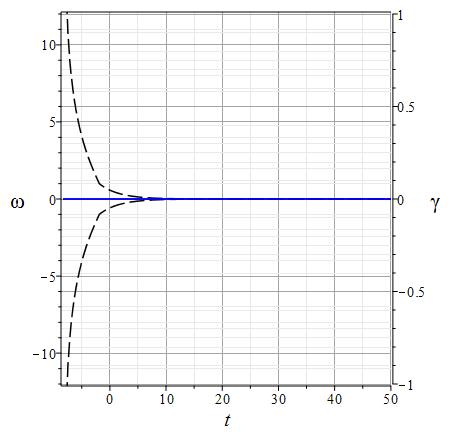}{6}{\label{fig6}Frequency evolution of $u^\pm_+(t)$ modes for a model with parameters \eqref{F_param1}, $n=1$.}
\Fig{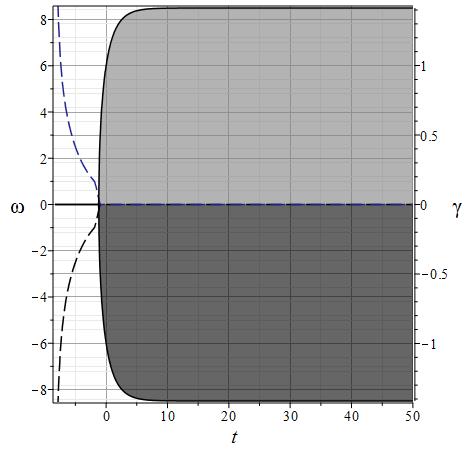}{6}{\label{fig7}Frequency evolution of $u^\pm_-(t)$ modes for a model with parameters \eqref{F_param1}, $n=1$.}

Further, as the wave number of perturbations $n$ increases, the growth of perturbations begins at a later time. On Fig. \ref{fig10} the dependence of the oscillation increment on $n$ is shown.\footnote{Figure from the work of the Author \cite{Yu_Unst_GC22_2}.} As can be seen from the figure, the beginning of the instability phase, $t_n$, approximately obeys the law:
\[t_n\backsimeq 8 \lg n.\]
\begin{Mark}\hspace{-6pt}\textbf{.} \label{mark6}
When switching to a similar model with real fundamental parameters ($k\sim 10^4\div10^5$), the corresponding time scale in this formula and on the graph in Fig. \ref{fig8} should be stretched by a factor of $k$, separating the distances between plots by about $10^5\div 10^6 t_{pl}$ .
\end{Mark}
\Fig{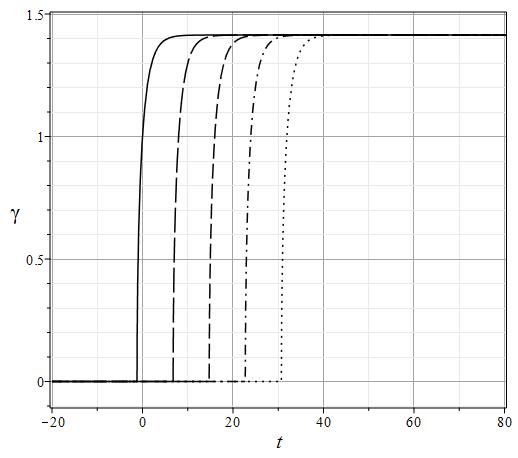}{6}{\label{fig8}Dependence of the evolution of the perturbation amplitude increment $u^-_-$ on the wave number $n$ for a model with parameters $[[1,1,1,10^{-5},0.1],10^{-5}]$, $n=1$.
Solid line -- n=1; long-dashed -- n=10; dashed -- n=100; dash-dotted -- n=1000; dashed -- n=10000.}
\section{SSBH  formation}
\subsection{Mechanism of formation of spherical perturbations}
As a result of the broadband scalar-gravitational instability, longitudinal perturbations in the cos\-mo\-lo\-gical medium can reach the nonlinear stage exponentially. How can flat perturbations \eqref{petrurb} be transformed into spherically symmetric masses and then into black holes? First, due to the macroscopic isotropy of the Universe, perturbations \eqref{petrurb} must also be distributed isotropically in directions, i.e., must be averaged by some spherically symmetric in the space $\mathbb{N}_3=\{\mathbf{n}\}$ distribution function $F(\mathbf{n})$ over two-dimensional angles on the sphere, as well as over the length of the wave vector $|\mathbf{n}|$, i.e., over the spectrum of perturbations\footnote{See see \cite{Yu_TMF_20}.} for details. In the standard theory of the formation of the large-scale structure of the Universe (see, for example, \cite{Zeld}), flat perturbations (the so-called ``pancakes'') ultimately survive, firstly, due to the weak growth of longitudinal perturbations in the fluid and the long time their exit to the nonlinear stage of shock waves, and, secondly, due to the survival of the perturbation mode with the highest amplitude of one random direction $\mathbf{n}$ in dusty matter. It is the dustiness factor of matter that leads to the formation of pancakes.

In the case of scalar-gravitational instability, the main component of matter is the scalar field, the dynamics of which is fundamentally different from the dynamics of non-interacting dust particles. Therefore, in the case of the development of scalar-gravitational instability, it is precisely spherically-symmetric mass distributions that form, under suitable conditions, into black holes. The following three factors are decisive in the formation of spherical disturbances from flat ones: 1. the isotropic nature of the disturbances; 2. dependence of the beginning of the instability phase of the perturbation mode on the wave number; 3. exponentially fast growth of perturbations due to which they rapidly become non-linear. Due to the isotropy of the distribution of perturbations, all perturbations with the same wave number begin to grow simultaneously, regardless of the direction of the wave vector. But in this case, perturbations with smaller values $n\sim1$ grow earlier. Due to their rapid growth and transition to the non-linear stage, it is they who survive, forming spherically symmetrical objects. The non-linear nature of perturbations under suitable conditions should lead to the formation of black holes in the future. The nature of the scalar-gravitational instability lies in a nonlinear combination of the features of the interparticle interaction of scalar-charged particles and the gravitational interaction \cite{Yu_Unst_GC23_2}.
\subsection{Evolution of localized spherically\newline symmetric perturbations}
To substantiate the applicability of the previous results to the case of spherically symmetric per\-tur\-ba\-tions in \cite{YuTMF_23}, the problem of the evolution of localized spherically symmetric perturbations that arose as a result of \emph{redistribution} of Friedmann's matter inside the sphere was solved. To describe small gravitational perturbations $\nu(r,t)$ and $\lambda(r,t)$, the metric in isotropic spherical coordinates\footnote{see, for example, \cite{Land_Field}} is used, which admits a continuous transition to the metric Friedman \eqref{ds_0} \cite{YuPhys1_08}:%
\begin{eqnarray}
\label{metric_pert}
\!\!\!ds^2=\mathrm{e}^{\nu}dt^2-a^2\mathrm{e}^{\lambda}[dr^2+r^2(d\vartheta^2+\sin^2\vartheta d\varphi^2)].
\end{eqnarray}
The localization of perturbations means that at some time $t=0$ the perturbations of the metric and the scalar field, together with their derivatives, vanish at the boundaries of the sphere of radius $r_0$
\begin{eqnarray}\label{bound_0}
\nu(r_0,0)=0; \quad {\displaystyle \left.\frac{d\nu(r,0)}{dr}\right|_{r=r_0}=0};\nonumber \\
\delta\Phi(r_0,0)=0; \quad {\displaystyle  \left.\frac{d\delta\Phi(r,0)}{dr}\right|_{r=r_0}=0}.
\end{eqnarray}
With the help of methods for extracting particle-like solutions and separating variables\footnote{For various aspects of this issue, see \cite{YuPhys1_08} -- \cite{YuPhys3_08}.} in \cite{YuTMF_23} it is shown, firstly, that near the singularity the metric \eqref{metric_pert} has the Schwarzschild asymptotics
{\small
\begin{eqnarray}\label{shvarc}
ds^2\backsimeq \biggl(1-\frac{2m(t)}{a(t)r}\biggr)dt^2 -\frac{dr^2+r^2(d\vartheta^2+\sin^2\vartheta d\varphi^2)}{\displaystyle 1-\frac{2m(t)}{a(t)r}}\nonumber
\end{eqnarray}
}
\noindent and, secondly, that in the process of cosmological evolution, the perturbation localization radius $r_0$ remains constant ($r_0=\mathrm{Const}$), so the physical perturbation localization radius $R_m$ grows with the expansion of the Universe:
\begin{equation}\label{R=ar0}
R_m=a(t)r_0.
\end{equation}
Due to the boundary conditions \eqref{bound_0} outside the sphere of radius $R(t)$ \eqref{R=ar0} matter <<does not know>> about the existence of a black hole, while the matter inside this sphere is forever gravitationally bound to it and gradually stretches black hole, increasing its mass. In this case, the problem of the evolution of localized perturbations is reduced to the problem of solving a chain of coupled ordinary linear differential equations, including an autonomous subsystem with respect to the singular mass $m(t)$ and charge $q(t)$. Third, numerical integration shows that, as a result of the evolution of spherical perturbations, the singular mass $m(t)$ can grow exponentially rapidly and reach the values \eqref{M_nc}.

\subsection{Evolution of the effective mass of the perturbed region}
The effective energy density of the cosmological system, $\mathcal{E}_{eff}$, is determined from the energy integral \eqref{Surf_Einst1_0} as follows (see \cite{TMF_21}):
\begin{equation}\label{E_eff}
\mathcal{E}_{eff}(t)=3H^2(t).
\end{equation}
According to \eqref{R=ar0} \emph{the effective mass of the perturbation region} $M(n,t)$ is defined as the total energy - the mass enclosed in a sphere of radius equal to the length of the perturbation $\ell_n(t)=a(t)/ n\equiv \mathrm{e}^{\xi(t)} /n$:
\begin{equation}\label{M_n}
m(n,t)=\frac{4\pi}{3}\lambda^3(t)\mathcal{E}_{eff}(t)=\frac{4\pi}{n^3}H^2(t)\mathrm{e}^{3\xi(t)}.
\end{equation}
Assuming $n=1$ in \eqref{M_n}, we introduce:
\begin{eqnarray}\label{M0}
m_0(t)\equiv m(1,t)=4\pi H^2(t)\mathrm{e}^{3\xi(t)}.
\end{eqnarray}
At the inflation stage ($H=H_0=\mathrm{Const}$, $\Omega=1$), according to \eqref{M_n}, the effective perturbation mass grows exponentially
\begin{equation}\label{M_n-H0}
m(n,t)=\frac{4\pi}{n^3}H^2_0\mathrm{e}^{3H_0 t}.
\end{equation}

Due to the property \textbf{\ref{stat3}}, the stage of inflationary expansion in the cosmological model $\Mod{M}$ can be reached twice: in the attractive foci $M^\pm_0$ \eqref{M_0} of the vacuum field model $\Mod{ M_0}$
and at its unstable singular points $M^\pm_\pm$ \eqref{M_pm} (\cite{arXiv_23}). Because \eqref{lambda0->Lambda} for $m_s\not\equiv0$ always
\begin{equation}\label{Lambda_0>Lambda}
\Lambda_0>\Lambda,
\end{equation}
then in the early stages the cosmological model first passes through the unstable singular point $M^\pm_\pm$ (stage \textbf{I}), from which it rolls down to the stable singular point $M^\pm_0$ from $\Phi\to0$ ( stage \textbf{II}). Thus, at the stages of inflation, according to \eqref{M_n-H0}, the effective mass of the perturbation harmonic $n$ evolves according to the law:
\begin{eqnarray}\label{M(y)_1}
\mathbf{1}: & \displaystyle m(n,t)=\frac{4\pi}{3n^3}\Lambda_0\mathrm{e}^{\sqrt{3\Lambda_0} t};\\
\label{M(y)_2}
\mathbf{2}: & \displaystyle m(n,t)=\frac{4\pi}{3n^3}\Lambda\mathrm{e}^{\sqrt{3\Lambda} t}.
\end{eqnarray}

At the time $\tau_g$, at which the wavelength of the perturbation (effective radius) becomes greater than its gravitational radius, the gravitational collapse of the spherical harmonic of the perturbation is possible. Thus, according to \eqref{M0} this point in time is determined by the equation:
\begin{equation}\label{Eq_tau_g}
\frac{8\pi}{n^2}H^2(\tau_g)\mathrm{e}^{2\xi(\tau_g)}=1.
\end{equation}
At the inflation stage, the equation \eqref{Eq_tau_g} has a solution:
\begin{equation}\label{tau_g}
\tau_g=\frac{1}{H_0}\ln\biggl(\frac{n}{\sqrt{8\pi}H_0}\biggr).
\end{equation}
The mass of the newborn black hole $m_g$ is obtained by substituting $t=\tau_g$ into \eqref{M0}
\begin{equation}\label{m_g}
m(n,\tau_g)=m_g.
\end{equation}
At the inflation stage, this gives:
\begin{equation}\label{M_g}
m_g=\frac{1}{4\sqrt{2\pi}}\frac{1}{H_0}
\end{equation}
is the mass of the newborn black hole does not depend on the wavenumber $n$ of the perturbation.

However, according to \eqref{tau_g}, the black hole birth time grows logarithmically with wavenumber $n$, so in perturbation modes with larger $n$ values, black holes are born later. From \eqref{M_g}
it follows that the masses of produced black holes, firstly, are larger at the \textbf{2} stage, and, moreover, larger for models with smaller values of the cosmological constant $\Lambda$. Secondly, taking into account the law of similarity transformation \eqref{Sol_tilde}, we find from \eqref{tau_g} and \eqref{M_g} expressions for the birth time and mass of a black hole in such a model:
\begin{eqnarray}\label{tilde_tau_g,m^0_bh}
\tilde{\tau}_g=k\tau_g;&
\tilde{m}_g=k m_g.
\end{eqnarray}
Therefore, in cosmological models with smaller values of fundamental parameters, black holes are born later, but at the same time - with larger masses.

\subsection{Influence of evaporation on the\newline formation of black holes}
Let us find out the influence of the Hawkin evaporation of black holes on the process of evolution of the black hole mass under study in a gravitational-scalar perturbation with wavenumber $n$. The rate of photon evaporation of a black hole with mass $m$ is described by the expression (see, for example, \cite{Wald}, \cite{DeWitt})
\begin{eqnarray}\label{dm/dt_Hawking}
\!\!\!\frac{dm}{dt}= -\frac{9}{5120\pi}\frac{\hbar c^4}{G^2m^2}\Rightarrow \frac{dm}{dt}=-\frac{9}{51 20\pi m^2}.
\end{eqnarray}
Calculating the time derivative of \eqref{M_n} and compiling the mass balance equation, we obtain a first-order inhomogeneous differential equation with respect to the mass of a black hole that appears in the perturbation harmonic with wavenumber $n$ as a result of the competition of collapse and evaporation processes \cite{Yu_Unst_GC22_3}:
\begin{eqnarray}\label{dM/dt_Balance}
\frac{dm}{dt}+\frac{9}{5120\pi}\frac{1}{m^2}=4\pi H^3(1+2\Omega)\frac{\mathrm{e}^{3\xi}}{n^3},
\end{eqnarray}
which must be solved with the initial condition \eqref{m_g}.

In order for the mass of a black hole to increase in the process of evolution, it is necessary that it increases at the time of the birth of the black hole $\tau_g$, i.e., according to the equation \eqref{dM/dt_Balance}, the following condition is satisfied:
\[4\pi H^3(\tau_g)(1+2\Omega(\tau_g))\frac{\mathrm{e}^{3\xi(\tau_g)}}{n^3}>\frac{9}{5120\pi}\frac{1}{m^2_g}.\]
Taking into account \eqref{M_n-H0} and \eqref{M_g} at the inflation stage ($\Omega=1$), this leads to the condition:
\begin{eqnarray}\label{dm/dt>0}
\!\!\!m^3_g>\frac{3}{5120\pi H_0}\Rightarrow H_0<\left(\frac{40}{3\sqrt{2\pi}}\right)^{\frac{1}{2}}\approx 2.306.
\end{eqnarray}
Note that, firstly, this condition, like the expression for the mass of a newborn black hole \eqref{M_g}, does not depend on the perturbation wavelength, but is completely determined only by the value of the Hubble parameter at the time of birth
of a black hole\footnote{However, according to \eqref{tau_g}, the moment of birth itself depends on the wave number.}, secondly, this condition is quite mild, since all the models we consider are characterized by much lower values of the Hubble parameter. Thus, the processes of evaporation of a black hole can only very slightly affect the process of its formation.
\begin{Mark} \label{mark7}\hskip -6pt\textbf{.}
Note that, according to \eqref{dm/dt_Hawking}, when the fundamental constants $c,G,\hbar$ are invariant with respect to the similarity transformation, the evaporating mass $m$ must be transformed according to the law $\tilde{m}=k^{1/3 }m$. But this law will contradict the transformation law of the right side of the \eqref{dM/dt_Balance} equation, which is invariant with respect to the similarity transformation. However, there is no contradiction here with the similarity properties of the studied cosmological model, since the process of evaporation of black holes is quantum and is external in relation to the studied classical model. So, the process of evaporation of black holes breaks the symmetry of similarity.
\end{Mark}

\section{Numerical modelling of SSBH formation}
In this section, as a basic one, we will study the $\Mod{M}$ model with realistic parameter values
\begin{equation}\label{Par_real}
\mathbf{P_2}: =\bigl[\bigl[10^{-8},10^{-4},10^{-9},10^{-3}\bigr],3\cdot10^{-11}\bigr],
\end{equation}
corresponding in order of magnitude to nonrelativistic fermions with a mass on the order of $10^{15}$ Gev in a field theory model of the SU(5) type. Moreover, according to \eqref{lambda0->Lambda} $\Lambda_0=2.53\cdot 10^{-9}$.
Sometimes, in order to find out the dependence of the model on the parameters, we will change one of the parameters of the model, keeping the values of the rest.

The dynamic system of this model corresponds to singular points \eqref{M_0} -- \eqref{M_pm}
\begin{eqnarray}\label{M_0_real}
M^\pm_0=\bigl[\pm3.163\cdot10^{-6},0,0\bigr];\\
\label{M_pm_real}
M^\pm_\pm=\bigl[\pm2.904\cdot10^{-5},\pm1,0\bigr],
\end{eqnarray}
the eigenvalues of the characteristic matrix \eqref{eigen_val_0} -- \eqref{eigen_val_pm} in which are:
\begin{eqnarray}\label{eigen_val_0_real}
M^\pm_0:& \lambda=\mp4.744\cdot10^{-6} \mp i\cdot 9.99\cdot10^{-5};\\
\label{eigen_val_pm_real}
M^\pm_\pm:& \lambda=[\mp 1.92\cdot10^{-4}, \pm1.044\cdot10^{-4}].
\end{eqnarray}
Thus, the points $M^\pm_0$ are attracting foci, the points $M^\pm_\pm$ are saddle foci.
\subsection{Basic functions of the model}
In the \eqref{Par_real} model, the singularity \eqref{t_0} is reached at time $t_0\approx -8.3281\cdot10^4$. On Fig. \ref{fig9} -- \ref{fig10} the early stages of the evolution of the cosmological model near this singularity are shown, corresponding to the total ultrarelativistic equation of state, i.e., the value of the cosmological acceleration $\Omega=-1$ (see also \cite{ Ignat_GC21}, where an analytical study of the behavior of the model near the singularity was carried out).

Next, in Fig. \ref{fig11} and \ref{fig12} show the large-scale evolution of scale functions and scalar potential.

\Fig{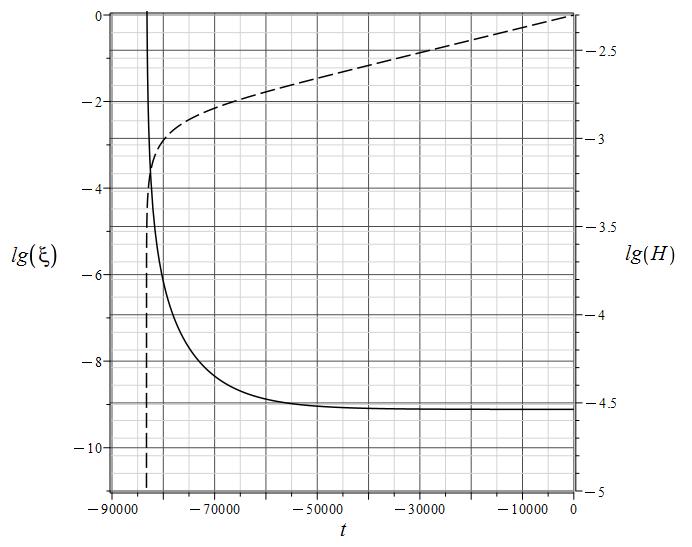}{6}{\label{fig9}Evolution of the scale functions $\xi(t)$ (dashed line) and $H(t)$ (solid line) near the singularity for a model with parameters \eqref{Par_real}.}
As can be seen from the graphs in Fig. \ref{fig11} -- \ref{fig12}, the cosmological model rolls from the unstable $M^+_+(H=1.92\cdot10^{-4},\Phi=1,Z=0)$ state to the stable $ M^+_0(H=4.744\cdot10^{-6},\Phi=0,Z=0)$, and the transition from the first inflation regime (1) to the second (2) occurs at time $t_1\approx3. 2\cdot10^5$ almost instantly.
\Fig{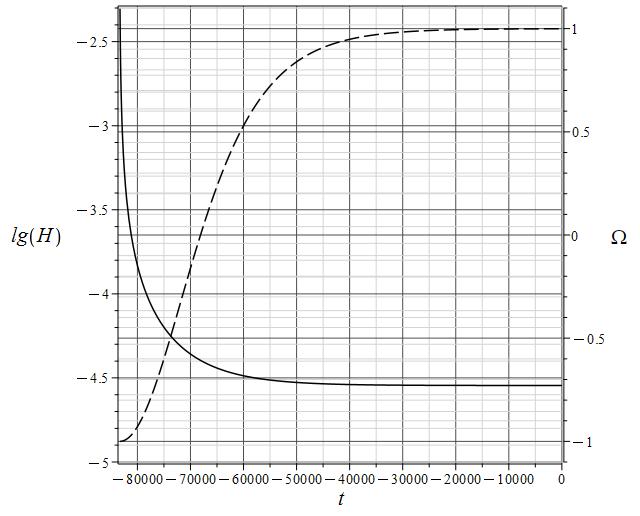}{6}{\label{fig10}Evolution of the scaling functions $H(t)$ (solid line) and $\Omega(t)$ (dashed line) near the singularity for a model with parameters \eqref{Par_real}.}
Thus, in the studied cosmological model, three stages of evolution are clearly distinguished:\\[12pt]
\noindent Stage \textbf{1}: $t=-8.3281\cdot10^4\div -4\cdot10^4$ -- ultrarelativistic stage ($\Omega=-1,\; \kappa=\frac{1}{ 3}$), stage duration $\Delta \tau_1\sim 4\cdot 10^4$;\\
\noindent Stage \textbf{2}: $t=-4\cdot10^4\cdot10^4\div 3.2\cdot10^5$ -- first inflation ($\Omega=1,\; \kappa=-1$, $H_I=2.904\cdot10^{-5},\ \Phi=1$), stage duration $\Delta \tau_{2}\sim 4\cdot 5.6\cdot10^5$;\\
\noindent Stage \textbf{3}: $t=3.2\cdot10^5\div +\infty$ -- second inflation ($\Omega=1,\; \kappa=-1$, $H_{II}=3.163 \cdot10^{-6},\ \Phi=0$), stage duration $\Delta \tau_{3}=\infty$.\\[12pt]

\Fig{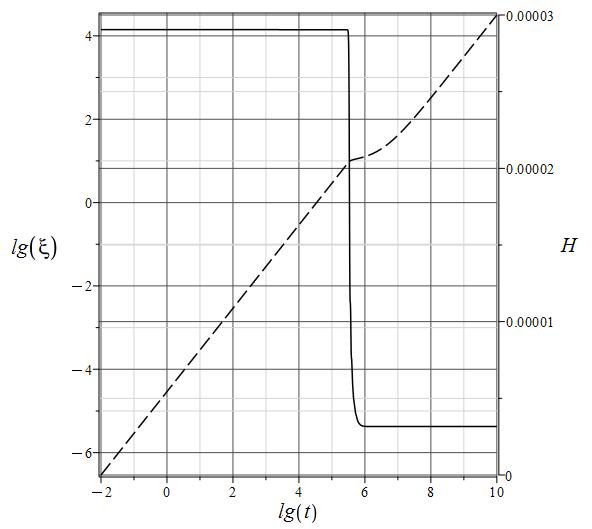}{6}{\label{fig11}Evolution of the scale functions $\xi(t)$ (dashed line) and $H(t)$ (solid line) on a large scale for a model with parameters \eqref {Par_real}.}
\Fig{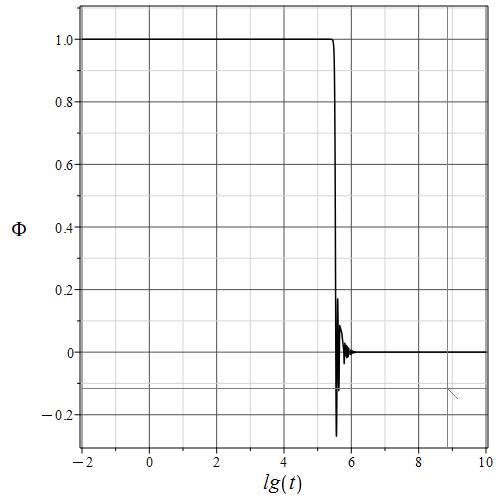}{6}{\label{fig12}Evolution of the scalar potential $\Phi(t)$ for a model with parameters \eqref{Par_real}.}
\subsection{Masses and birth times of black holes}
Let us study the dependence of the mass $m_g$ \eqref{m_g} and the time $\tau_g$ \eqref{tau_g} of black hole birth in the \eqref{Par_real} model on the wavenumber of perturbations. On Fig. \ref{fig13} dependency plots are presented
the mass of the newborn black hole and the time of its birth on the wave number $n$. As we noted above, the mass of a newborn black hole does not depend on the wave number, and the time of its birth is proportional to the logarithm of the wave number.
The only exceptions are perturbation modes with small values of the wave number $n\lesssim 1$.

Note that in this model, for $n\gtrsim1$, black holes are born with mass $m_g\sim3\cdot 10^4$ at times $\tau_g\sim10^6\div7\cdot10^6$. These times correspond to the inflationary stage \textbf{2}.

\Fig{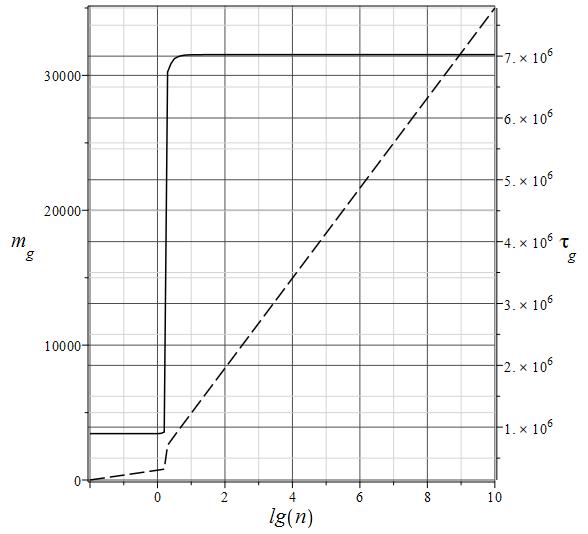}{6}{\label{fig13}Dependence of the black hole mass $m_g$ (solid line) and birth time $\tau_g$ (dashed line) on the wavenumber $n$ for the model with parameters \eqref {Par_real}.}
\subsection{Evolution of the effective perturbation mass}
On Fig. \ref{fig14} -- \ref{fig16} shows the results of numerical simulation of the evolution of the effective per\-tur\-ba\-tion mass $m(n,t)$ according to the formula \eqref{M_n}. The gray horizontal stripe in the following figures marks the area required masses SSBH \eqref{M_nc}. In particular, in Fig. \ref{fig14} the influence of the wavenumber $n$ of the gravitational perturbation on the evolution of the effective mass in the cosmological model with parameters \eqref{Par_real} is shown.
\Fig{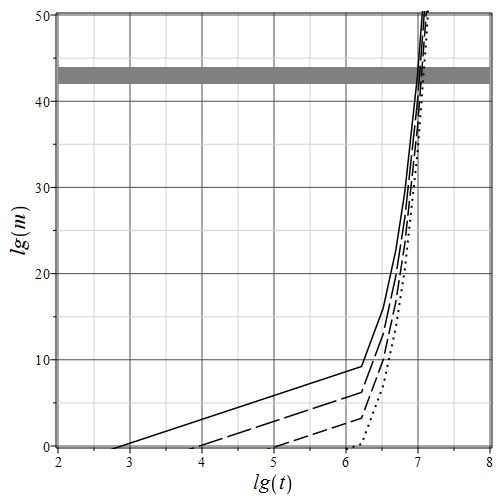}{7}{\label{fig14}Evolution of the effective perturbation mass $m(n,t)$ as a function of the wave number $n$ for a model with parameters \eqref{Par_real}. From left to right: $n=1$, $10$, $100$, $1000$.}
On Fig. \ref{fig15} -- \ref{fig16} the evolution of the perturbation effective mass depends on the cosmological constant $\Lambda$ and the scalar charge $e$. In this case, parameters close to $\mathbf{P_2}$ \eqref{Par_real} are chosen when all parameters except one are fixed.
\begin{eqnarray}\label{Par_Lambda}
\mathbf{P_\Lambda}: =\bigl[\bigl[10^{-8},10^{-4},10^{-9},10^{-3}\bigr],\Lambda\bigr],\\
\label{Par_e}
\mathbf{P_e}: =\bigl[\bigl[10^{-8},10^{-4},e,10^{-3}\bigr],3\cdot10^{-11}\bigr],
\end{eqnarray}

Note that on all graphs \ref{fig14} -- \ref{fig16} presented in this section, a strong increase in the effective perturbation mass begins from the time $t\backsimeq 10^6$, i.e., according to the graphs in Fig. \ref{fig11} and \ref{fig12} -- immediately after the transition to the inflationary stage \textbf{2}, corresponding to the stable singular point of the dynamical system.
\Fig{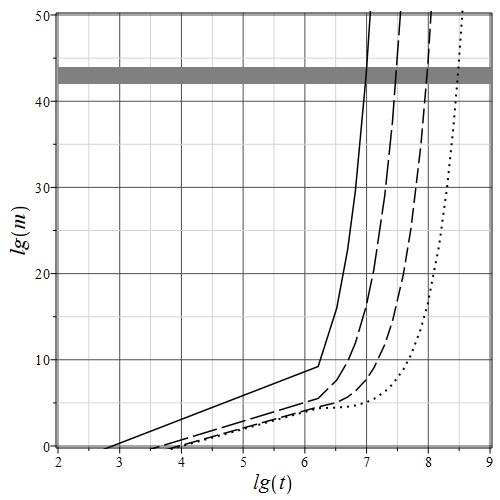}{7}{\label{fig15}Evolution of the perturbation effective mass $m(n,t)$ as a function of the cosmological constant $\Lambda$ for a model with parameters \eqref{Par_Lambda}. From left to right: $\Lambda=3\cdot10^{-11}$, $3\cdot10^{-12}$, $3\cdot10^{-13}$, $3\cdot10^{-14}$.}
\Fig{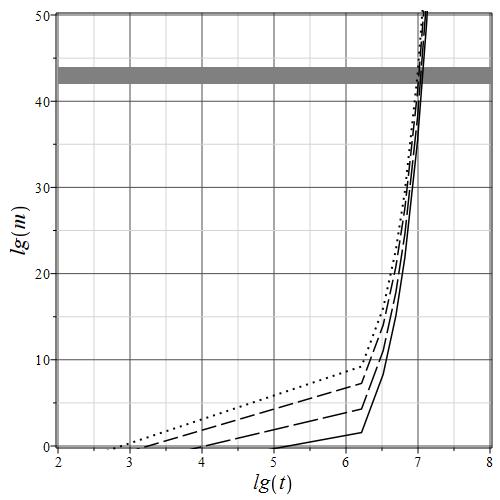}{7}{\label{fig16}Evolution of the perturbation effective mass $m(n,t)$ depending on the value of the scalar charge $e$ for the model with parameters \eqref{Par_e}. From right to left: $e=\cdot10^{-3}$, $\cdot10^{-5}$, $\cdot10^{-7}$, $\cdot10^{-9}$. \eqref{Par_e}.}
\subsection{Evolution of black holes\newline with evaporation}
In this section, we will demonstrate the results of numerical simulation of the process of evolution of black holes formed at time $\tau_g(n)$ based on the differential equation \eqref{dM/dt_Balance}. In our model $H\ll1$, therefore, the condition \eqref{dm/dt>0} is met with a huge margin, i.e., evaporation processes cannot interrupt the opposite process of growth in the mass of the formed black hole. However, this process needs to be studied more carefully. On Fig. \ref{fig17} shows a large-scale picture of the evolution of the mass of a newborn black hole depending on the wave number $n$ of the perturbation. As can be seen from this figure, firstly, as before, a rapid increase in mass begins immediately after the transition of inflation to the \textbf{2} stage. Secondly, at the \textbf{1} stage and at the later stages of the \textbf{2} stage, the evolution of the black hole mass depends very weakly on the wave number of the collapsed perturbation.
\Fig{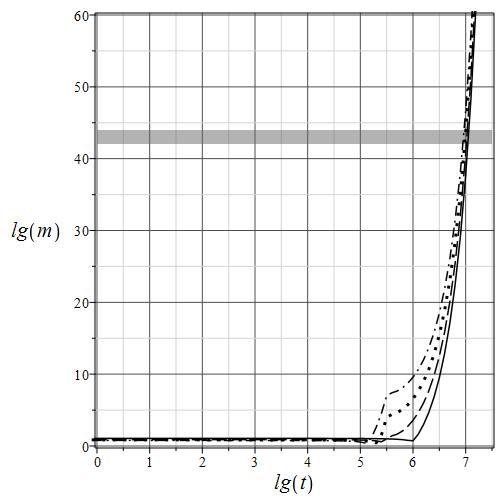}{7}{\label{fig17}Evolution of the black hole mass $m(n,t)$ depending on the wave number $n$ for the model with parameters \eqref{Par_real}. From left to right: $n=0.1$, $1$, $10$, $100$. }
\Fig{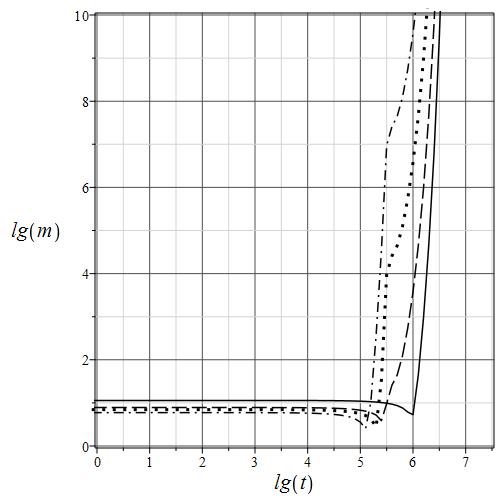}{7}{\label{fig18}Detailed evolution of the black hole mass $m(n,t)$ depending on the wave number $n$ for the model with parameters \eqref{Par_real}. From left to right: $n=0.1$, $1$, $10$, $100$. }
Significant differences are observed just at the point of change in the inflation regime, which can be seen in the detailed picture in Fig. \ref{fig18}.

It can be seen that near the transition point, the black hole mass losses due to evaporation become significant, but still they do not have time to stop the mass growth process. We cannot completely rule out the possibility of the opposite picture in models with other fundamental parameters.

\section*{Conclusion}
Summing up the intermediate results of this part of the article, we note, firstly, that the use of the similarity transformation in combination with numerical methods has made it possible to sig\-ni\-fi\-cantly expand the range of parameters of cos\-mo\-lo\-gical models for studying the process of formation of supermassive black holes in the early Universe and, in a number of cases, to obtain analytical formulas that describe these processes . In particular, studies of processes with realistic fundamental parameters of interactions, predicted by the existing field theory models of elementary particles, turned out to be accessible. In this work, we used the parameters of the SU(5) field model. If use parameters of the so-called standard model (SM) in relation to the similarity transformation, we should choose something like the following list:
\[\mathbf{P_{sm}}=\bigl[\bigl[10^{-30},10^{-15},10^{-15},10^{-9}\bigl],3\cdot10^{-33}\bigl],\]
which is obtained from the list $\mathbf{P_2}$ \eqref{Par_real} by similarity transformation with coefficient $k=10^{11}$. In such a model, according to \eqref{tilde_tau_g,m^0_bh}, we get the characteristic time of the birth of a black hole $\tau_g\sim 10^{17}$, and the mass of the newborn black hole is $m_g\sim 3\cdot10^{14}$. In this case, the characteristic time of the beginning of the rapid growth of the black hole mass (the beginning of the 2nd inflation) also shifts to a value on the order of $10^{17}$, and the required SSBH mass range $m_{ssbh}$ \eqref{M_nc} is reached in times of the order of $10^{18}$, which is still very far from the critical time for observations of 1 mlr. years. Thus, the timely formation of SSBH occurs in the standard model as well.

\emph{The questions to be answered are the following:}
\begin{enumerate}
\item According to the above results, the process of increasing the mass of SSBH does not stop when the required mass \eqref{M_nc} is reached, but continues indefinitely. Now we need to find a mechanism to stop this process.
\item What is the large-scale structure of the\\ Universe after the completion of this process, what is the fate of the matter that fell into the sphere of influence of SSBH?
\end{enumerate}

The second part of the article will be devoted to the answers to these questions.\\[12pt]

\noindent \textbf{Thanks}\\
The author is grateful to the participants of the seminar of the Department of Theory of Relativity and Gravity of Kazan University for the useful discussion of some aspects of the work. The author is especially grateful to professors S.V. Sushkov and A.B. Balakin for very useful comments and discussion of the features of modern modifications of the theory of gravity as applied to cosmology and astrophysics. The Author is also grateful to the participants of the 5th International Winter School-Seminar <<Peter's Readings-2022>> for the fruitful discussion of a number of issues raised in the article, and especially to Professors K.A. Bronnikov, M.O. Katanaev and B. Sakha.\\[12pt]

\noindent \textbf{Founding}\\
The work was carried out at the expense of a subsidy allocated as part of the state support of the Kazan (Volga Region) Federal University in order to increase its competitiveness among the world's leading scientific and educational centers.


\end{document}